\documentclass[fleqn,twoside]{article}
\usepackage[headings]{espcrc2}
\usepackage{epsfig,amsfonts}
\usepackage{graphicx}
\usepackage{dcolumn}
\usepackage{bm}
\DeclareMathSymbol{\lesssim}      {\mathrel}{AMSa}{"2E}
\DeclareMathSymbol{\gtrsim}       {\mathrel}{AMSa}{"26}
\def\be{\begin{equation}}
\def\ee{\end{equation}}
\def\bc{\begin{center}}
\def\ec{\end{center}}
\def\bea{\begin{eqnarray}}
\def\eea{\end{eqnarray}}

\def\nn{\nonumber}
\title{Two-loop QED Corrections to Bhabha Scattering}

\author{R.~Bonciani\address[IFIC]{Departament de F\'{\i}sica Te\`orica, 
        IFIC, CSIC -- Universitat de 
        Val\`encia, E-46071 Val\`encia,
        Spain}%
        \thanks{This work was supported by the European Union under the 
	contract HPRN-CT2002-00311 (EURIDICE) and by MCYT (Spain) under 
	Grant FPA2004-00996, by Generalitat Valenciana (Grants GRUPOS03/013 
	and GV05/015).} and 
	A.~Ferroglia\address[Freiburg]{Physikalisches Institut, 
	Albert-Ludwigs-Universit\"at Freiburg, D-79104 Freiburg, 
	Germany}}

\begin{document}

\begin{abstract}
Recent developments in the calculation of the NNLO corrections 
to the Bhabha scattering differential cross section in pure QED
are briefly reviewed and discussed. 

\vspace{1pc}
\end{abstract}

\maketitle

\section{Status of the NNLO corrections}

Bhabha scattering, $e^{+}e^{-} \to e^{+}e^{-}$, is a crucial process in the
phenomenology of particle physics. Its relevance 
is mainly due to the fact that it is the process employed to determine the 
luminosity ${\mathcal L}$ at $e^{+}e^{-}$ colliders: in fact, 
${\mathcal L} = {\mathcal N}_{Bhabha} / \sigma_{th}$,
where ${\mathcal N}_{Bhabha}$ is the rate of Bhabha events and $\sigma_{th}$ is
the Bhabha scattering cross section calculated from theory.

Two kinematic regions are of special interest for the  luminosity measurements,
since in these regions the Bhabha scattering cross section is comparatively
large and QED dominated.
At colliders operating at c. m. energies of  ${\mathcal O}  (100 \,
\mbox{GeV})$, the relevant kinematic region is the one in which the angle
between the outgoing particles and the beam line is of few degrees.  This, for
instance, was the case at LEP, where the luminometers were located between
50 and 100  mrad, and it will be also the case at the future ILC (luminometers
between 25 and  80 mrad \cite{Moenig}). 
At machines operating at c. m. energies of the order of $1-10
\, \mbox{GeV}$, the region of interest is instead the one in 
which the scattering angle is large; as an example, at KLOE,  
the luminosity measurement involves Bhabha scattering events that take place 
at angles between $55^{\circ}$ and $125^{\circ}$ \cite{Denig}.
Moreover, the large angle Bhabha scattering  will be employed at the ILC 
in order to study the beam beam effects that lead to a non monochromatic
luminosity spectrum \cite{non-mon}. 

Experimentally, Bhabha scattering is measurable with a very high accuracy. 
At LEP, the experimental error on the luminosity measurement was 
$4 \cdot 10^{-4}$ \cite{LEP}. At ILC it is expected to be of the same order 
of magnitude or better (the goal of the TESLA forward calorimeter collaboration 
is to reach, in the first year of run, an experimental error of 
$1 \cdot 10^{-4}$ \cite{TESLA}). 
This remarkable accuracy requires, as a counterpart, an equally precise theoretical
calculation of the Bhabha scattering cross section, in order to keep the
luminosity error small. Therefore, radiative corrections to the basic process
have to be under control.

In order to match the detector geometry and experimental cuts of any 
particular machine, a Monte Carlo  event generator is needed.
In the recent past, several groups have been working on Monte Carlo generators
for Bhabha scattering, in both the large-angle and small-angle 
kinematic regions. 
The LEP theoretical simulations of Bhabha events were based on BHLUMI
\cite{BHLUMI}, whose theoretical error, mainly due to missing higher 
order corrections, is estimated to be $4.5 \cdot 10^{-4}$ (see for instance 
\cite{Jadach}). The KLOE collaboration employs the Monte Carlo 
event generators BABAYAGA \cite{BABAYAGA} and BHAGENF \cite{BHAGENF}, 
which have an estimated theoretical error of $5 \cdot 10^{-3}$;
 within the error claimed, they are 
in agreement with each other. Moreover, BHWIDE \cite{BHWIDE} and MCGPJ 
\cite{MCGPJ} provided valuable checks.
All the mentioned Monte Carlo programs for Bhabha scattering employ the 
mass of the electron as a cut-off for collinear divergences; this is to be 
taken into account  when calculating NLO ($\mathcal{O}(\alpha^3)$) and
 NNLO ($\mathcal{O}(\alpha^4)$) corrections to the cross
section.

The complete $\mathcal{O}(\alpha^3)$ corrections  to Bhabha scattering, in the
full Electroweak Standard Model, have been known  for a long time
\cite{Bhabha1loop}. The corrections of $\mathcal{O}(\alpha^4)$ to the
differential cross section  in the Standard Model  are not yet known. In recent
years, several papers were devoted to the study of NNLO corrections in pure
QED. In \cite{russians,Fadin:1993ha,Arbuzov:1995vj} the second  order radiative
corrections, both virtual and real, enhanced by factors of  $\ln^n(s/m^2)$
(with $n=1,2$, $s$ the c. m. energy squared, and $m$ the mass of the electron) where
studied.  The complete set of these corrections was finally  obtained in
\cite{Bas}.  This was achieved by employing the QED virtual corrections for
massless  electron and positrons of \cite{Bern} and  the results of \cite{bmr},
as well as by using  the known structure of the IR poles in dimensional
regularization \cite{Catani}.  In \cite{Penin:2005kf}, the complete set of
photonic $\mathcal{O}(\alpha^4)$  corrections to the differential cross section
that are not suppressed by positive powers  of the ratio $m^2/s$ was
calculated. The virtual corrections of  $\mathcal{O}(\alpha^4)$ involving a
closed fermion loop, together with  the corresponding soft-photon emission
corrections, were obtained in  \cite{us}; no mass expansion or approximation
was employed, and the result  retains the full dependence on the electron mass
$m$. The calculation was performed by means of the
Laporta-Remiddi algorithm \cite{Lap} which takes advantage of the 
integration by parts \cite{IBP} and Lorentz-invariance \cite{LI} identities 
in order to reduce the problem to the calculation of a small set of master 
integrals. The master integrals were calculated using the differential 
equations method \cite{DiffEq}; their expression in terms of 
harmonic polylogarithms \cite{HPLs} is given in \cite{MIsus}. 
Several papers deal with the unapproximated calculation of the master integrals
necessary for the evaluation of the photonic NNLO corrections \cite{MIs}; in
particular, the reduction to master integrals of these corrections is complete,
and only the  master integrals related to the two-loop boxes have not yet
been all evaluated. The status of the calculation of these master integrals
is discussed in \cite{TR}. 
In \cite{bf}, finally, the unapproximated calculation of the photonic vertex
contributions to the $\mathcal{O}(\alpha^4)$ Bhabha scattering cross section, 
as well as the calculation of the subset of radiative corrections due to the interference of
one-loop diagrams (already considered in \cite{Fleischer:2002wa}) was completed.
A few papers discuss the leading NNLO weak corrections to Bhabha scattering
cross section \cite{EW}. 

In summary, the complete NNLO corrections to Bhabha scattering in the
full Standard Model are  still far from being completely known. 
Concerning the pure QED contributions, from a phenomenological point of view,  
all the numerically relevant corrections to the NNLO differential cross section 
are known, with the only exception of the ones arising from 
the production of soft pairs, for which only the terms enhanced by 
$\ln^n(s/m^2)$ with $n=1,2$ are present. 
If we consider instead the exact fixed order calculation, the situation is 
less satisfactory. The  unapproximated two-loop QED photonic box contributions
are  still missing. Moreover, 
it would be interesting to evaluate the corrections arising from diagrams with
closed loops of heavier fermions (like muons or taus).

\section{The Small Mass Limit \label{small}}

\begin{figure}
\bc
\begin{picture}(0,0)%
\includegraphics[scale=0.6]{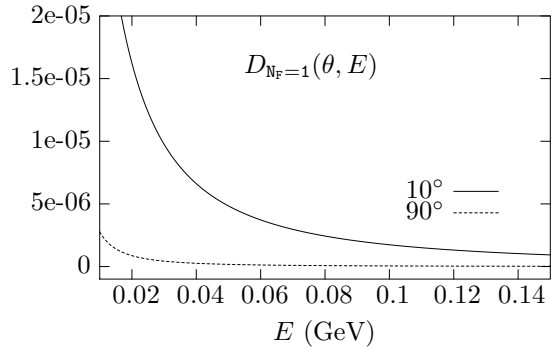}%
\end{picture}%
\setlength{\unitlength}{0.0120bp}%
\begin{picture}(18000,10800)(0,0)%
\put(2750,2374){\makebox(0,0)[r]{\strut{} 0}}%
\put(2750,4343){\makebox(0,0)[r]{\strut{} 5e-06}}%
\put(2750,6312){\makebox(0,0)[r]{\strut{} 1e-05}}%
\put(2750,8281){\makebox(0,0)[r]{\strut{} 1.5e-05}}%
\put(2750,10250){\makebox(0,0)[r]{\strut{} 2e-05}}%
\put(4036,1430){\makebox(0,0){\strut{} 0.02}}%
\put(6057,1430){\makebox(0,0){\strut{} 0.04}}%
\put(8079,1430){\makebox(0,0){\strut{} 0.06}}%
\put(10100,1430){\makebox(0,0){\strut{} 0.08}}%
\put(12121,1430){\makebox(0,0){\strut{} 0.1}}%
\put(14143,1430){\makebox(0,0){\strut{} 0.12}}%
\put(16164,1430){\makebox(0,0){\strut{} 0.14}}%
\put(10100,275){\makebox(0,0){\strut{}$E$ (GeV)}}%
\put(7573,8675){\makebox(0,0)[l]{\strut{}$D_{\tt N_F=1}(\theta, E)$}}%
\put(13868,4737){\makebox(0,0)[r]{\strut{}$10^{\circ}$}}%
\put(13868,4062){\makebox(0,0)[r]{\strut{}$90^{\circ}$}}%
\end{picture}%
\vspace*{-9mm}
\caption{\it{$D_{\tt N_F=1}$ as a function of the beam energy, for 
 $\theta = 10^{\circ}$ (solid line) and $\theta = 90^{\circ}$ (dashed line). 
 The soft-photon energy cut-off is set equal to $E$}.}
\label{deltaNF}
\ec
\end{figure}

At the colliders mentioned above, the mass of the electron is very small in 
comparison with the c. m. energy. Therefore, it is reasonable to 
expect that it can be safely ignored except in the terms where it acts as a 
cut-off for collinear divergencies, as it was done in obtaining the results 
of \cite{Penin:2005kf}. For the set of corrections obtained in \cite{us} and 
\cite{bf}, for which unapproximated analytic results are available, 
it is possible to determine the numerical relevance of the terms suppressed 
by positive powers of the ratio $m^2/s$ as a function of the beam energy.
\begin{figure}
\bc
\begin{picture}(0,0)%
\includegraphics[scale=0.6]{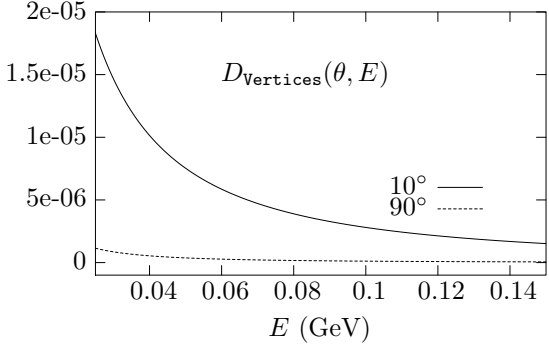}%
\end{picture}%
\setlength{\unitlength}{0.0120bp}%
\begin{picture}(18000,10800)(0,0)%
\put(2750,2373){\makebox(0,0)[r]{\strut{} 0}}%
\put(2750,4342){\makebox(0,0)[r]{\strut{} 5e-06}}%
\put(2750,6311){\makebox(0,0)[r]{\strut{} 1e-05}}%
\put(2750,8281){\makebox(0,0)[r]{\strut{} 1.5e-05}}%
\put(2750,10250){\makebox(0,0)[r]{\strut{} 2e-05}}%
\put(4723,1429){\makebox(0,0){\strut{} 0.04}}%
\put(6987,1429){\makebox(0,0){\strut{} 0.06}}%
\put(9251,1429){\makebox(0,0){\strut{} 0.08}}%
\put(11515,1429){\makebox(0,0){\strut{} 0.1}}%
\put(13779,1429){\makebox(0,0){\strut{} 0.12}}%
\put(16043,1429){\makebox(0,0){\strut{} 0.14}}%
\put(10100,275){\makebox(0,0){\strut{}$E$ (GeV)}}%
\put(6987,8281){\makebox(0,0)[l]{\strut{}$D_{\tt Vertices}(\theta, E) $}}%
\put(13504,4736){\makebox(0,0)[r]{\strut{}$10^{\circ}$}}%
\put(13504,4061){\makebox(0,0)[r]{\strut{}$90^{\circ}$}}%
\end{picture}%
\vspace*{-9mm}
\caption{\it{$D_{\tt Vertices}$ as a function of the beam energy, for 
$\theta = 10^{\circ}$ (solid line) and $\theta = 90^{\circ}$ (dashed line). 
The soft-photon energy cut-off is set equal to $E$}.}
\label{deltaV}
\ec
\end{figure}

This analysis is performed in \cite{bf}. The three contributions taken
into account are the one arising from graphs with a  closed fermion loop
($N_F=1$), the photonic corrections involving at least a vertex graph
({\tt Vertices}), and the interference of one-loop box  diagrams ({\tt Box Box}). For each
contribution, the quantity
\bea
D_{i} \! \! &=& \! \! \left(\frac{\alpha}{\pi}\right)^2 \left|\left( 
\frac{d \sigma_2^{(i)}}{d \Omega} -
\left. \frac{d \sigma_2^{(i)}}{d \Omega} \right|_L 
\right) \right| \nn\\
\! \! & & \! \! \times \left(\frac{d \sigma_0}{d \Omega} +
\left(\frac{\alpha}{\pi}\right) \frac{d \sigma_1}{d \Omega}
\right)^{-1}
\, ,
\label{Ds}
\eea
with $i=(N_F=1), \mbox{\tt Vertices}, \mbox{\tt Box Box}$, is plotted as a 
function of the beam energy $E$. In Eq.~(\ref{Ds}), $d \sigma_2^{(i)}/d \Omega$
is the unapproximated $\mathcal{O}(\alpha^4)$ correction to the cross 
section taken into account.
It includes the contribution of  the virtual diagrams and the one of
the corresponding diagrams with the 
emission of up to two soft photons (with energy smaller than the cut-off 
$\omega$). $d \sigma_2^{(i)}/d \Omega |_L$ is the same quantity as 
$d \sigma_2^{(i)}/d \Omega$, aside for the fact that the terms proportional to
positive powers of the ratio $m^2/s$ are neglected.

\begin{figure}
\bc
\begin{picture}(0,0)%
\includegraphics[scale=0.6]{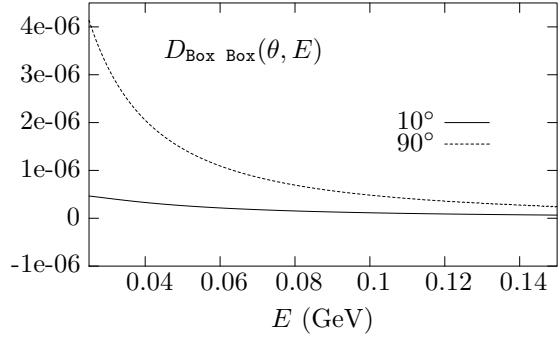}%
\end{picture}%
\setlength{\unitlength}{0.0120bp}%
\begin{picture}(18000,10800)(0,0)%
\put(2200,1979){\makebox(0,0)[r]{\strut{}-1e-06}}%
\put(2200,3483){\makebox(0,0)[r]{\strut{} 0}}%
\put(2200,4987){\makebox(0,0)[r]{\strut{} 1e-06}}%
\put(2200,6490){\makebox(0,0)[r]{\strut{} 2e-06}}%
\put(2200,7994){\makebox(0,0)[r]{\strut{} 3e-06}}%
\put(2200,9498){\makebox(0,0)[r]{\strut{} 4e-06}}%
\put(4239,1429){\makebox(0,0){\strut{} 0.04}}%
\put(6591,1429){\makebox(0,0){\strut{} 0.06}}%
\put(8943,1429){\makebox(0,0){\strut{} 0.08}}%
\put(11295,1429){\makebox(0,0){\strut{} 0.1}}%
\put(13647,1429){\makebox(0,0){\strut{} 0.12}}%
\put(15999,1429){\makebox(0,0){\strut{} 0.14}}%
\put(9825,275){\makebox(0,0){\strut{}$E$ (GeV)}}%
\put(4827,8746){\makebox(0,0)[l]{\strut{}$D_{\tt Box \,\,\, Box}(\theta, E)$}}%
\put(13372,6490){\makebox(0,0)[r]{\strut{}$10^{\circ}$}}%
\put(13372,5815){\makebox(0,0)[r]{\strut{}$90^{\circ}$}}%
\end{picture}%
\vspace*{-9mm}
\caption{\it{$D_{\tt Box \,\,\, Box}$ as a function of the energy, 
for $\theta = 10^{\circ}$ (solid line) and $\theta = 90^{\circ}$ 
(dashed line). The soft-photon energy cut-off is set equal to 
$E$}.}
\label{deltaBB} 
\ec
\end{figure}

In Figs. \ref{deltaNF}, \ref{deltaV} and \ref{deltaBB} the functions $D_i$,
evaluated for two different sample angles ($10^{\circ}$ and $90^{\circ}$), 
are shown. It is clear that the approximation in which terms proportional to
positive powers of the ratio $m^2/s$ are neglected  is extremely good already 
at energies that are significantly smaller than the ones encountered in  
$e^{+}e^{-}$ experiments.

The NNLO corrections arising from the two-loop photonic boxes are not taken
into account in this analysis, since unapproximated result for that set of
corrections are not available. 
Nevertheless, it is reasonable to expect in the $m^2/s \to 0$ limit a behavior
similar to the one of the other NNLO corrections.

\section{Results}

\begin{figure}
\bc
\begin{picture}(0,0)%
\includegraphics[width=76mm,height=60mm]{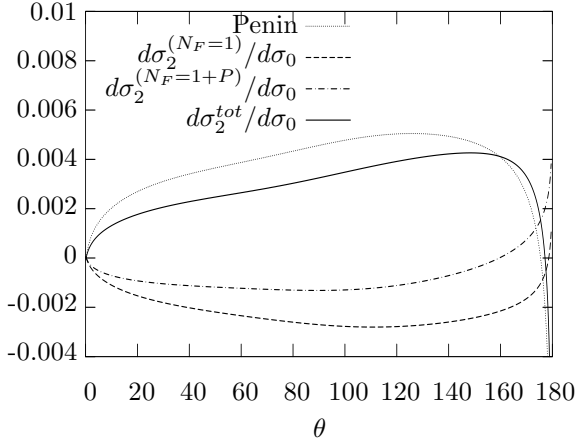}%
\end{picture}%
\setlength{\unitlength}{0.0120bp}%
\begin{picture}(17995,13800)(0,0)%
\put(2200,2700){\makebox(0,0)[r]{\strut{}-0.004}}%
\put(2200,4230){\makebox(0,0)[r]{\strut{}-0.002}}%
\put(2200,5760){\makebox(0,0)[r]{\strut{} 0}}%
\put(2200,7290){\makebox(0,0)[r]{\strut{} 0.002}}%
\put(2200,8820){\makebox(0,0)[r]{\strut{} 0.004}}%
\put(2200,10350){\makebox(0,0)[r]{\strut{} 0.006}}%
\put(2200,11880){\makebox(0,0)[r]{\strut{} 0.008}}%
\put(2200,13410){\makebox(0,0)[r]{\strut{} 0.01}}%
\put(2475,1429){\makebox(0,0){\strut{} 0}}%
\put(4100,1429){\makebox(0,0){\strut{} 20}}%
\put(5726,1429){\makebox(0,0){\strut{} 40}}%
\put(7351,1429){\makebox(0,0){\strut{} 60}}%
\put(8976,1429){\makebox(0,0){\strut{} 80}}%
\put(10602,1429){\makebox(0,0){\strut{} 100}}%
\put(12226,1429){\makebox(0,0){\strut{} 120}}%
\put(13852,1429){\makebox(0,0){\strut{} 140}}%
\put(15478,1429){\makebox(0,0){\strut{} 160}}%
\put(17095,1429){\makebox(0,0){\strut{} 180}}%
\put(9825,275){\makebox(0,0){\strut{} $\theta$ }}%
\put(9142,12996){\makebox(0,0)[r]{\strut{}Penin}}%
\put(9142,12146){\makebox(0,0)[r]{\strut{}$d\sigma_{2}^{(N_{F}=1)}/ d\sigma_{0}$}}%
\put(9142,11100){\makebox(0,0)[r]{\strut{}$d\sigma_{2}^{(N_{F}=1+P)}/ d\sigma_{0}$}}%
\put(9142,10046){\makebox(0,0)[r]{\strut{}$d\sigma_{2}^{tot}/ d\sigma_{0}$}}%
\end{picture}%
\vspace*{-9mm}
\caption{\label{all} \it{Photonic, $N_F=1$, and total contributions to the 
cross section at order $\alpha^4$. The beam energy is chosen equal to 
$0.5$ {\rm GeV} and the soft-photon energy cut-off $\omega$ is set equal to
$E$. }}
\ec
\end{figure}

In Fig. \ref{all} all the QED contributions to the $\mathcal{O}(\alpha^4)$ 
Bhabha scattering cross section known at the moment 
are plotted as a function of the scattering angle. Terms suppressed by positive
powers of the ratio $m^2/s$ were neglected.

The dotted line represents the photonic corrections \cite{Bas,Penin:2005kf}. The 
corrections of ${\mathcal O}(\alpha^4 (N_F=1))$ \cite{us} (dashed line) have, 
for this choice of $\omega$ ($\omega = E$), an opposite sign with respect to 
the photonic ones. Moreover they include large terms 
proportional to $\ln^3(s/m^2)$ that cancel out once the contribution of the 
soft pair production is included. Concerning the latter, only the terms proportional 
to $\ln^n(s/m^2)$ ($n=1,2,3$) are known (see \cite{Arbuzov:1995vj}). The 
dashed-dotted line in the figure represents the sum of the 
${\mathcal O}(\alpha^4 (N_F=1))$ cross section with the known terms of the pair production
corrections\footnote{The pair production contribution depends upon 
a cut-off on the energy of the soft electron-positron pair. In the numerical 
evaluation of Fig.~\ref{all} we set
\be
\ln({\mathrm{D}}) = \frac{1}{2} \ln \left( \frac{4 \Omega^2}{s} \right) \, ,
\ee
with $\ln({\mathrm{D}})$ defined in \cite{Arbuzov:1995vj} and $\Omega$  
numerically equal to the soft-photon cut-off: $\Omega = \omega$.}. The solid 
line, finally, is the complete order $\alpha^4$ QED Bhabha scattering cross 
section, including photonic, $N_F=1$ and pair production contributions.

To conclude, we presented the status of the NNLO QED corrections to the Bhabha
scattering differential cross section, paying particular attention to the
phenomenological relevance of the terms that are suppressed by positive powers
of the ratio $m^2/s$. It turns out that, in view of the precision required by 
present and future experiments, the NNLO QED corrections can not be neglected,
and should be included in the Monte Carlo event generators.

\section{Acknowledgment}

The authors wish to thank the participants of RADCOR05 in particular 
A. Penin, C. Carloni Calame, S. Jadach and L. Trentadue for useful 
discussions. Special thanks to the organizers.

\end{document}